\documentclass{aa}
\usepackage{graphicx,natbib}

\def\arcsec{$^{\prime\prime\,}$}

\def\*{$^{*}$}

\def\apjl{ApJL}
\def\apj{ApJ}
\def\apjs{ApJ Suppl.Ser.}

\begin{document}

\sloppypar
 
\title{Discovery of X-ray pulsations from IGR J16320$-$4751 = \mbox {AX
J1631.9$-$4752}}
 
\author{A.Lutovinov \inst{1,2}, J.Rodriguez\inst{3,4},
M.Revnivtsev\inst{2,1}, P.Shtykovskiy\inst{1,2}}

\offprints{lutovinov@hea.iki.rssi.ru} 

\institute{Space Research Institute, Russian Academy of Sciences,
              Profsoyuznaya 84/32, 117810 Moscow, Russia
\and
Max-Planck-Institute f\"ur Astrophysik, Karl-Schwarzschild-Str. 1, D-85740
Garching bei M\"unchen, Germany
\and
CEA Saclay, DSM/DAPNIA/SAp (CNRS FRE 2591), 91191 
Gif-sur-Yvette Cedex, France
\and
Integral Science Data Center, Chemin d'Ecogia, 16, 1290 Versoix, Switzerland
}

\date{Received /Accepted}

        \authorrunning{Lutovinov et al.}  \titlerunning{Discovery of
        X-ray pulsations from IGR J16320$-$4751 = AX J1631.9$-$4752}
        
   \abstract{We report a discovery of strong modulations of the X-ray
   flux detected from IGR J16320$-$4751 = AX J1631.9$-$4752 with a
   period of $P\sim1300$ sec. We reanalyzed the data of an XMM-Newton
   ToO performed soon after the discovery of the source by INTEGRAL
   and found the modulation at a period of $P=1309\pm40$ sec with a
   high significance. Modulations of the source flux with two possible
   periods of $\sim1300$ and $\sim1500$ sec were identified in the
   ASCA archival data. It is very likely that the modulation can be
   interpreted as X-ray pulsations, favouring a pulsar as the compact
   object in IGR/AX J16320$-$4752. Thus for the moment this source
   became the fourth source from a new class of highly absorbed binary
   systems for which the pulsations are observed.
    }

   \maketitle
%

\section{Introduction}

For the moment about a dozen new hard X-ray sources were discovered
(or rediscovered) with the INTEGRAL observatory in the direction to
the Norma Galactic spiral arm. Most of them as believed are members of
high mass X-ray binaries with early type primary stars. From three of
such objects (IGR J16358$-$4726, AX J163904$-$4642, IGR J16465$-$4507)
X-ray pulsations were detected \cite{pat04,wal04,lut04}.

The transient hard X-ray source IGR J16320$-$4751 was discovered on Feb 1,
2003 with the INTEGRAL observatory during ToO observations of 4U1630-47
\cite{tom03}. The source demonstrated a significant high energy variability
on a time scale of several thousand of seconds \cite{tom03,fosch04}. Its
position was coincident with the one reported by Sugizaki et al. (2001) for
\mbox{AX J1631.9$-$4752}, which was observed previously by ASCA (therefore 
below we will call this source as IGR/AX J16320$-$4752). The follow-up
observation of IGR/AX J16320$-$4752 with XMM-Newton on March 4, 2003
confirmed a complex behaviour of the source lightcurve and
demonstrated flare-like events with durations of \mbox{$\sim1$ ksec}
without significant variations in the hardness
\cite{rod03}. The total exposure of this observation was about $\sim25$
ksec, but due to soft proton flares only $\sim$4.9 ksec of the data were
used for the analysis. Rodriguez et al. (2003) searched for the pulsations
in the power spectrum of the source, but due to the short good time interval
selected could not obtain strong constraints.

In this letter we reanalyze the XMM-Newton data of March 2003 and the
archival data of ASCA observations of Sep. 1997 with a special
attention to long period variations and show that the source
demonstrates clear pulsations with a period about $\sim 1.3$ ksec. The
source was also observed simultaneously with INTEGRAL and XMM-Newton
observatories in August 2004. Detailed spectral and temporal analysis
of these two data sets will be presented in a forthcoming paper. We
note, however, that a preliminary quick look analysis of those data
confirmed the presence of pulsations in IGR/AX 16320$-$4751
(L. Foschini, private comm.). These pulsations on the $\sim1.3$ ksec
time scale could not have detected before by the INTEGRAL observatory
due to the source faintness and the large time bin size ($\sim15-20$
ksec) chosen for the lightcurve building \cite{fosch04}.

\section{Data reduction}

IGR/AX J16320$-$4752 was observed with the XMM-Newton observatory on
March 4, 2003. In our analysis we used data of the EPIC-PN camera,
which operated in the Large Window science mode, thus offering a time
resolution of 48 msec.  The data were processed with the Science
Analysis System (SAS) v6.0.0.  As was mentioned above during the
previous scientific spectral and timing analysis \cite{rod03} the
significant part of data was excluded from the analysis due to proton
flares.  Such a filtering reduces the length of useful observing time
to $\sim5$ ksec and therefore makes searches for long periodic
variations almost impossible. Since the proton flares originate from
the interaction of the soft protons in the Earth's magnetosphere with
the telescope, their timing behavior is supposed to have no periodic
structure.  Therefore, for our purpose of a periodic signal search,
this step could be omitted and we applied practically no filtering to
the data.  We used only the simplest criteria, i.e.  selected events
with pattern 0-4 (singles and doubles) and flag \#XMMEA\_EP.  A
barycentric correction was applied to all selected counts. The
lightcurve was extracted from the 40\arcsec circle around the source
with its total duration being $\sim22$ ksec.

\begin{figure}
\includegraphics[width=\columnwidth,bb=30 176 590 720,clip]{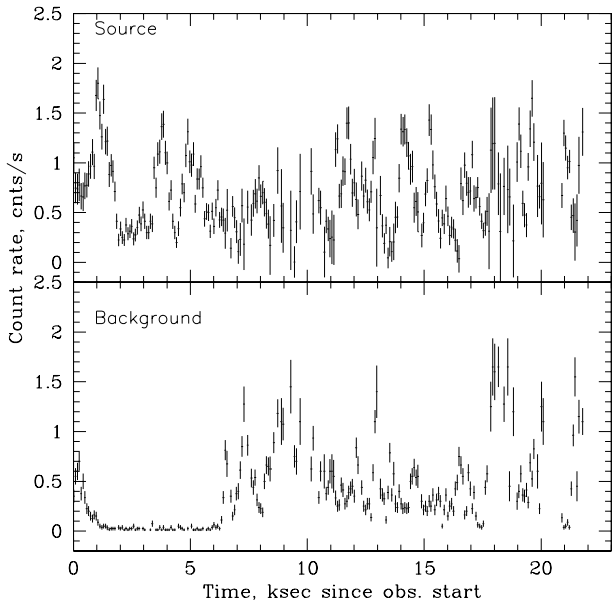}
\caption{Lightcurve of the source (upper panel) and the background (lower
  panel) during March 4, 2004 observation of IGR/AX J16320$-$4752 with
  XMM-Newton}
\label{lcs}
\end{figure}

\begin{figure}
\includegraphics[width=\columnwidth,bb=35 185 565 650,clip]{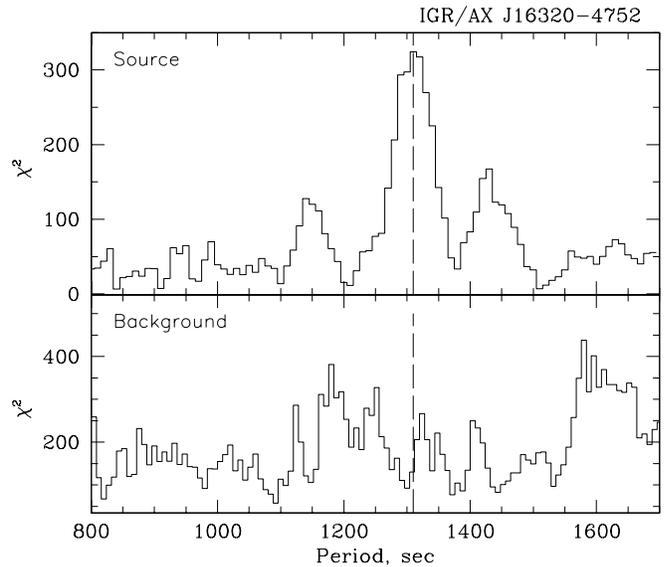}
\caption{$\chi2$ distribution obtained while folding the source and the
  background lightcurves during March 4, 2004 observation of IGR/AX
  J16320$-$4752 with XMM-Newton. It is seen that the background
  lightcurves does not demonstrate a coherent signal with the period
  of $\sim$1300 sec.}
\label{chi2_bkg}
\end{figure}

The region around the source was observed several times by ASCA, but
here we will analyze only data with the largest exposure time
($\sim13.7$ ksec) obtained on Sep. 4, 1997. Another observation of the
source with ASCA (Sep.3, 1994) when it was significantly detected has
only $\sim$3 ksec exposure time and is not suitable for our study.  In
order to increase the statistics we analyzed joint data of GIS2 and
GIS3 telescopes. The source was out of the field of view of both SIS
detectors.  We analyzed data with the help of standard tasks of
LHEASOFT/FTOOLS 5.2 package in accordance with ASCA Guest Observer
Facility recommendations.

\section{Results}

The lightcurves of the source (background subtracted) and the
background obtained with XMM-Newton on March 4, 2004 in the energy
band 1-10 keV are presented in Fig. \ref{lcs}. It is seen that the
background is quite strong during most of the observation.  However,
in spite of this the background subtracted lightcurve gives an
impression about possible periodic variations of the source flux.
Search for pulsations with the epoch folding technique in these two
lightcurves gives the results presented in Fig. \ref{chi2_bkg}. Very
significant peak at $P\sim1300$ sec is clearly seen for the source
lightcurve, while the background curve does not show any peak at this
period.  Therefore we can be sure that the periodic variability can
not be caused by the background variations. The statistical
significance of pulsations in the lightcurve of IGR/AX J16320$-$4752
can be estimated either with the help of standard formulas (e.g. Leahy
et al. 1983) or directly from the data after the construction of the
statistical distribution of obtained $\chi^2$ values for trial periods
apart from the pulse period. The statistical significance of the
detection of pulsations with the XMM-Newton data exceeds 10$\sigma$,
and can be considered as unambiguous.

\begin{figure}
\includegraphics[width=\columnwidth,bb=34 176 576 750,clip]{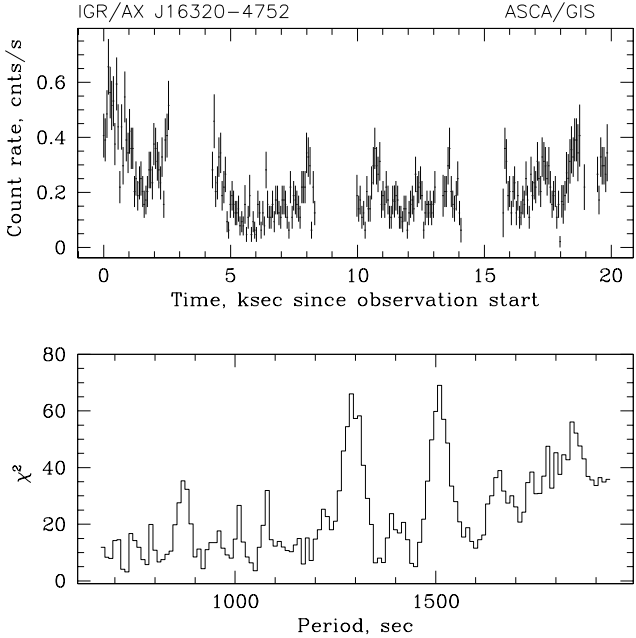}
\caption{Lightcurve of IGR/AX J16320$-$4752 obtained by ASCA/GIS (upper panel) 
and $\chi2$ distribution obtained while folding this lightcurve for
various trial periods (lower panel).}
\label{chi2}
\end{figure}

The lightcurve of the source observed by ASCA during its Sep 4, 1997
observation is presented in Fig.\ref{chi2} (upper panel).  The
dependence of $\chi^2$ value on the value of trial periods used for
the epoch folding of this lightcurve is presented in Fig.\ref{chi2}
(lower panel).  As it is seen from the figure there are two clear
candidates for the coherent variations of the X-ray flux, around 1300
sec and around 1500 sec. Two possible periods can appear as a result
of gaps in the ASCA data, due to the short $\sim 90$ minutes orbital
period, and a significant ``red'' noise in the source variability (we
can clearly see the long term decline of the source flux during first
few ksec).  However, performed Monte-Carlo simulations of the
lightcurve with gaps and red noise of different forms have not allowed
us to solidly distinguish the real period from the possible ``alias''
period. Therefore it seems reasonable to consider below two possible
values of the source period in 1997.

The best fit period of the pulsations of IGR/AX J16320$-$4752 according to the
XMM-Newton data (March 2004) is $P=1309\pm40$ sec. The $\chi^2$-dependence
for the ASCA data demonstrates several peaks with two most prominent at
$P=1292\pm40$ sec and $P=1510\pm50$ sec.

\begin{figure}
\includegraphics[width=\columnwidth,bb=34 176 576 480,clip]{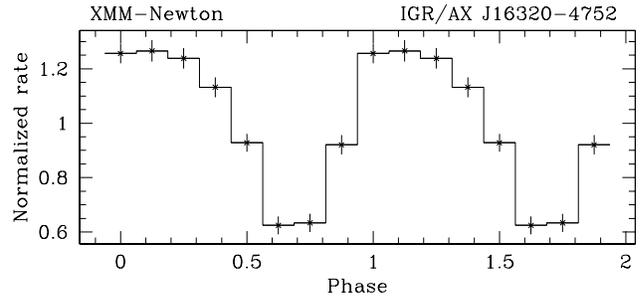}
\caption{Pulse profile of IGR/AX J16320$-$4752 obtained in the 1-10 keV
  energy band by folding of the XMM-Newton lightcurve with the period
  of 1309 sec.}
\label{folded}
\end{figure}

The lightcurve of IGR/AX J16320$-$4752 obtained with XMM-Newton
folded with the period of 1309 sec is presented in Fig.\ref{folded}.
Phases on this plots are arbitrary.  The pulse fraction (which is
defined as $P=(I_{\rm max}-I_{\rm min})/(I_{\rm max}+I_{\rm min})$,
where $I_{\rm max}$ and $I_{\rm min}$ are intensities at the maximum
and minimum of the pulse profile) are $26\pm3$\%.  We do not present
here the ASCA folded lightcurve because it is not clear what was the
pulsation period of the source at that time.

The time difference between observations of the source with ASCA and
XMM-Newton is relatively large, approximately 7 years, that leaves the
possibility for discussion of pulse period variations of IGR/AX
J16320$-$4752.  Accreting X-ray pulsars change their periods as a
result of a spin-up induced by the angular momentum gain, transported
to the neutron star by the accreting matter or a spin-down, caused by
the angular momentum loss e.g. due to the ''propeller'' effect.
Typical values of the accretion induced spin-up rate are approximately
in the range of $10^{-13}-10^{-11}$ Hz s$^{-1}$ depending on the
accretion rate in the binary system (see e.g. Bildsten et
al. 1997). If we assume that the real period of the pulsar in the
system IGR/AX J16320$-$4752 during the ASCA observation (year 1997)
was $\sim1500$ sec than we should conclude that the source have
demonstrated the average spin-up rate at the level of $\sim
5\times10^{-13}$ Hz s$^{-1}$ during next seven years. This value
corresponds to a source luminosity of $L\sim 2\times 10^{36}$ erg
s$^{-1}$ in the simple model of a magnetized (a magnetic moment
$\mu\sim10^{30}$ G cm$^3$) neutron star \cite{ghlam79}.  Alternatively
we can estimate the source luminosity from its spectrum.  The
absorption corrected flux of the source in the broad energy band
($\sim 1-100$ keV) during ASCA observations assuming the same spectral
shape at high energy as it was seen by INTEGRAL \cite{fosch04, lut04}
is approximately $\sim 5\times10^{-10}$ erg s$^{-1}$ cm$^{-2}$, which
corresponds to the source luminosity at the distance of 8 kpc
(assuming the source is close to the Norma spiral arm tangent) $L\sim
4\times 10^{36}$ erg s$^{-1}$, that is in an agreement with the above
estimations.

Naturally, we can not exclude that the real pulse period of the source
during ASCA observation was close to $1300$ sec. In this case periods
of a high accretion rate (spin-up) in the system should alternate with
periods of a spin down, similar to what is observed in other X-ray
pulsars (e. g. Lutovinov et al. 1994, Bildsten et al. 1997).

\section{Summary}

We report the discovery of X-ray pulsations from IGR/AX J16320$-$4752
with help of XMM-Newton and ASCA data. The study of the XMM-Newton
observation performed in 2003 allow us to unambiguously identify the
true period of pulsations $P=1309\pm40$ sec. This confirmed a
proposition of \cite{rod03} about the neutron star as a compact object
in this high-mass X-ray binary system. The result obtained from the
analysis of the ASCA data has not very high statistical significance
and have two almost equally significant pulse period candidates,
$\sim1300$ sec and $\sim 1500$ sec. If the latter value is the real
period of IGR/AX J16320$-$4752 during the ASCA observations in 1997,
then we can conclude that the source should have a relatively
persistent emission during last seven years providing the average
spin-up rate $\dot{\nu}\sim5\times10^{-13}$ Hz s$^{-1}$. We cannot
exclude the source to be truly persistent, but the sensitivity of
current all sky monitors is not enough to detect it in a such crowded
region, like a Norma arm. If the former value is the real pulse period
in 1997, then the source should alternate between high and low
accretion rate regimes.

\begin{acknowledgements}
Authors thank Luigi Foschini for the support of this work. Authors also
thank an anonymous referee for valuable comments and suggestions. AL and MR
acknowledge the support of RFFI grant 04$-$02$-$17276.  JR acknowledges
financial support from the French Space Agency (CNES).
\end{acknowledgements}


\begin{thebibliography}{}

\bibitem[Bildsten et al. 1997]{bild97} Bildsten,
  L., Chakrabarty, D., Chiu, J., et al. 1997, ApJS, 113, 367 

\bibitem[Ghosh \& Lamb 1979]{ghlam79}Ghosh P. \& Lamb F. 1979, ApJ, 234, 296

\bibitem[Foschini et al. 2004]{fosch04}  Foschini L., Tomsick J., Rodriguez
  J. et al. 2004, Proceeding of V INTEGRAL Workshop, Munich (ESA SP-552),
  247  

\bibitem[Leahy et al. 1983]{leahy83}Leahy, D.,
Darbro, W., Elsner, R., et al. 1983, \apj, 266, 160

\bibitem[Lutovinov et al. 1994]{lut94} Lutovinov,
A., Grebenev, S., Sunyaev, R., \& Pavlinsky, M. 1994, Astron. Lett., 20, 538

\bibitem[Lutovinov et al. 2004]{lut04} Lutovinov
  A., Revnivtsev M., Gilfanov M. et al. 2004, A\&A submitted,
  astro-ph/0411550 

\bibitem[Patel et al. 2004]{pat04} Patel, S.,
Kouveliotou, C., Tennant, A., et al.\ 2004, \apjl, 602, L45

\bibitem[Rodriguez et al. 2003]{rod03} Rodriguez,
  J., Tomsick, J.,  Foschini, L., et al. 2003, A\&A, 407, L41

\bibitem[Sugizaki et al. 2001]{sug01} Sugizaki, M., 
Mitsuda, K., Kaneda, H. et al.\ 2001, \apjs, 134, 77

\bibitem[Tomsick et al. 2003]{tom03}Tomsick, J., 
Lingenfelter, R., Walter, R., et al. 2003, IAUC, 8076


\bibitem[Walter 2004]{wal04} Walter, 
R.~\& INTEGRAL Survey Team 2004, AAS/High Energy Astrophysics Division, 8,  



\end{thebibliography}
\end{document}